\documentclass[twocolumn]{aastex63}
\submitjournal{ApJS}
\shorttitle{A reference transient dataset I: lightcurves}
\shortauthors{Neira et al.}

\usepackage{import}
\usepackage{hyperref}
\usepackage[T1]{fontenc}

\usepackage{graphicx}
\usepackage{amsmath}
\usepackage{amssymb}

\usepackage[utf8]{inputenc}
\usepackage{multirow}
\usepackage{float} 

\begin{document}
\title{MANTRA: A Machine Learning reference lightcurve dataset for astronomical transient event recognition}

\author{Mauricio Neira}
\affiliation{
Systems and Computing Engineering Department\\
Universidad de los Andes\\
Cra. 1 No. 18A-10\\
Bogot\'a, Colombia}

\author{Catalina G\'omez}
\affiliation{Departamento de Ingenier\'ia Biom\'edica\\
Universidad de los Andes\\ 
Cra. 1 No. 18A-10\\
Bogot\'a, Colombia\\}

\author{John F. Su\'arez-P\'erez}
\affiliation{Departamento de F\'isica\\
Universidad de los Andes\\
Cra. 1 No. 18A-10\\
Bogot\'a, Colombia}

\author{Diego A. G\'omez}
\affiliation{
Systems and Computing Engineering Department\\
Universidad de los Andes\\
Cra. 1 No. 18A-10\\
Bogot\'a, Colombia}

\author{Juan Pablo Reyes}
\affiliation{
Systems and Computing Engineering Department\\
Universidad de los Andes\\
Cra. 1 No. 18A-10\\
Bogot\'a, Colombia}

\author{Marcela Hern\'andez Hoyos}
\affiliation{
Systems and Computing Engineering Department\\
Universidad de los Andes\\
Cra. 1 No. 18A-10\\
Bogot\'a, Colombia}

\author{Pablo Arbel\'aez}
\affiliation{Departamento de Ingenier\'ia Biom\'edica\\
Universidad de los Andes\\ 
Cra. 1 No. 18A-10\\
Bogot\'a, Colombia\\}

\author{Jaime E. Forero-Romero}
\affiliation{Departamento de F\'isica\\
Universidad de los Andes\\
Cra. 1 No. 18A-10\\
Bogot\'a, Colombia}

\begin{abstract}
We introduce MANTRA, an annotated dataset of $4869$ transient and
$71207$ non-transient object lightcurves built from the Catalina Real
Time Transient Survey.
We provide public access to this dataset as a plain text file to facilitate
standardized quantitative comparison of astronomical transient event
recognition algorithms. 
Some of the classes included in the dataset are: supernovae, cataclysmic
variables, active galactic nuclei, high proper motion stars, blazars
and flares.
As an example of the tasks that can be performed on the dataset
we experiment with multiple
data pre-processing methods, feature selection techniques and popular
machine learning algorithms (Support Vector Machines, Random Forests
and Neural Networks).   
We assess quantitative performance in two classification tasks:
binary (transient/non-transient) and eight-class classification.   
The best performing algorithm in both tasks is the Random Forest Classifier.
It achieves an F1-score of $96.25\%$ in the binary
classification and $52.79\%$ in the eight-class classification.
For the eight-class classification, non-transients
($96.83\%$) is the class with the highest F1-score, while the lowest corresponds to high-proper-motion stars ($16.79\%$); for supernovae it achieves a value of $54.57\%$, close to the average
across classes. 
The next release of MANTRA includes images and benchmarks with deep
learning models.  
\end{abstract}

  
\section{Introduction}

Large scale automatic detection and classification of astronomical
transients is happening within surveys such as Pan-STARRS1
\citep{2004SPIE.5489...11K}, the Palomar Transient Factory
\citep{2009PASP..121.1395L},  the Catalina Real-Time Transient Survey
\citep{2009ApJ...696..870D}, the All-Sky Automated Survey for
SuperNovae \citep{2014ApJ...788...48S} and the Zwicky Transient
Factory \citep{2019PASP..131a8002B}.
Besides the large amount of data, transient
classification is hard because the data is usually heterogeneous,
unbalanced, sparse, unevenly sampled and with missing information.   

These two characteristics (size and heterogeneity) have
motivated the application of Machine Learning (ML) algorithms to face
this challenge.  
For instance, Random Forests, MultiLayer Perceptron and K-Nearest
Neighbours have been used on lightcurves to classify transients from
the Catalina Real Time Transient Survey \citep{disanto};  
convolutional neural networks have been used as 
input to automatic vetting algorithms (quick classification of bogus
vs. real transients) based on data from the SkyMapper Supernova and
Transient  Survey and the High cadence Transient Survey (HiTS)
\citep{1708.08947,1701.00458}.

The quality of the dataset is a necessary, although insufficient,
requirement for the success of these examples and of any other ML implementation.
New ML results usually come from groups internal to an observational
collaboration because they have the internal know-how (and, sometimes,
privileged access) to build training datasets.
Consequently, despite having a lot of data available in the public domain,
much of it is difficult to access.
This difference in data access makes it challenging for the broader
astronomical and machine learning communities
to rebuild a training dataset, perform comparisons with
published results and suggest new algorithms. 

Other collaborations have directly published large datasets of \textit{simulated transients}
hoping to trigger more involvement from the ML in astronomy
community at large to develop new classification algorithms \citep{2018arXiv181000001T}.
However, at the time of writing, no dataset for transients has been made accessible  to the public in the form of a catalogue based on real data.

To this end we compile and publish in easy-to-access files a dataset
that can be used to train and test different ML algorithms for
transient classification. 
We use public data from the Catalina Real-Time Transient Survey
(CRTS) \citep{1111.2566}, an astronomical survey searching transient
and highly variable objects as base for the dataset.
Effectively, we developed an ETL (Extract, Transform and Load) procedure to extract
the data from CRTS, which is designed for a sporadic lookup of a few objects, into a 
catalogue of thousands of objects that can be used to train ML algorithms and establish benchmarks.
Here, in the first paper, we present the lightcurve data.
In a second paper we will present a curated imaging dataset from the same survey.   

This paper is structured as follows.
In Section \ref{sec:data} we present the CRTS and the steps we follow
to build the dataset.
Then, in Section \ref{sec:repository} we describe its main features together
with the repository structure gathering the files and Python code to explore it.
In Section \ref{sec:ml_tests} we show how this dataset can be used to 
perform tests using ML methods following a similar approach as \cite{disanto},
and the experiments that we perform. 
We finalize in Section \ref{sec:conclusions} with a summary of the
main features of our dataset and the results of our experiments. 

\section{The lightcurve dataset} 
\label{sec:data}

We use public data from the Catalina Real-Time Transient Survey
(CRTS) \citep{2009ApJ...696..870D, 2011BASI...39..387M}, an astronomical survey searching transient
and highly variable objects.
The CRTS covered 33000 squared degrees of sky and took data since 2007. 
Three telescopes were used: Mt. Lemmon Survey (MLS), Catalina Sky 
Survey (CSS), and Siding Spring Survey (SSS). So far, CRTS has 
discovered more than $15000$ transient events.
We use data from the CSS telescope, which is an f/1.8 Schmidt
telescope located in the Santa Catalina Mountains in Arizona.
The telescope is equipped with a 111-megapixel  detector, and covered
4000 square degrees per night, with a limiting magnitude of 19.5 in
the V band.  

The web interface to access CRTS data has been primarily designed 
to query individual objects, not thousands as is our intention.
Our initial efforts to consolidate a lightcurve catalog used 
a variety of web scraping techniques.
A final consolidation of the catalog required the help of the
CRTS collaboration to dump the raw database files into a legacy webpage, as it proved
unfeasible to build the whole transient catalog through their
web interface.
However, we still keep the non-transients from the web scrapping extractions.

Putting together the lightcurves for MANTRA implies
cross-matching different files in the legacy CRTS webpage:
\url{http://nesssi.cacr.caltech.edu/DataRelease/CRTS-I_transients.html}. 
The photometry is stored in two different kinds of files: \verb"phot"
that come from the main photometry database and \verb"orphan" that
correspond to transients not associated with the 500 million sources
in the main photometry database.
There are also \verb"out" files that must be used to link transient
IDs to database IDs.

\begin{figure*}
\begin{center}
  \includegraphics[width=0.9\textwidth]{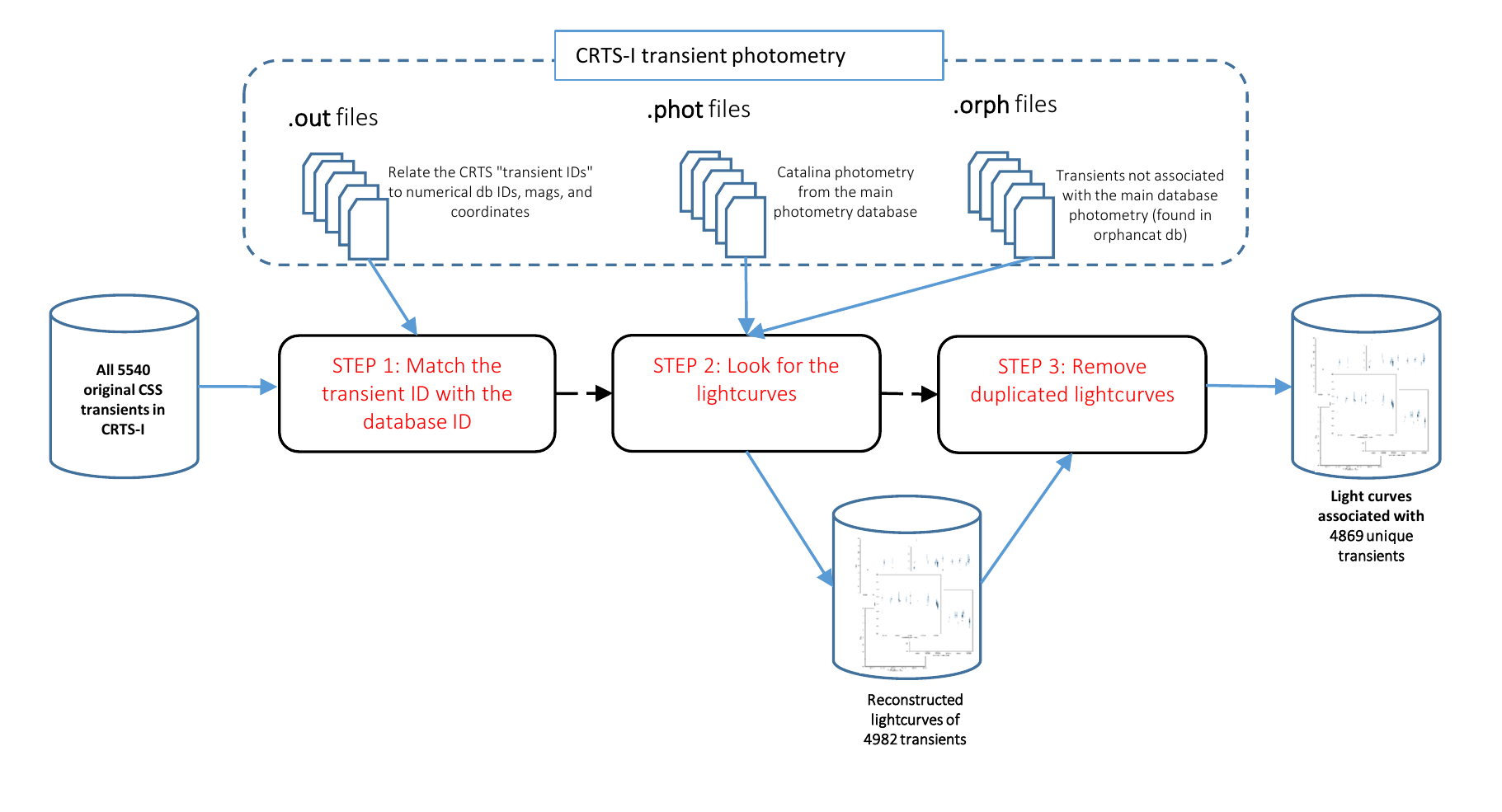}
\end{center}
  \caption{MANTRA Dataset Set Up: Lightcurve compilation for transient classes.}
  \label{fig:transients}
\end{figure*}

For each one of the 5540 transients reported and classified in the
archival webpage \url{http://nesssi.cacr.caltech.edu/catalina/All.arch.html} we use its transient IDs and its database IDs to look for the lightcurves in the
\texttt{phot} and \texttt{orphan} files. 
Only 4982 transients can be linked to available data to reconstruct
their lightcurves. 
Furthermore, some of these lightcurves are duplicated, i.e. they had
the same number of observations, Modified Julian Date (MJD) and magnitude measurement. 
We ignore the duplicates to end up with 4869 unique transients with an 
associated lightcurve. Figure \ref{fig:transients} summarizes this process. 

The CRTS dataset already provides a classification. 
The most numerous classes are: supernovae,
cataclysmic variable stars, blazars, flares, asteroids, active
galactic nuclei, and high-proper-motion stars (HPM). 
Though most objects in the transient object catalogue belong to a single class, 
there is some uncertainty in the categorization of some of them.  
In this case, an interrogation sign is used when a class is not clear 
e.g. SN? or sometimes multiple possible classes are found for a single 
event e.g. SN/CV.
Table \ref{table:top_classes} summarizes the number of objects in each class. 

\begin{figure*}
\begin{center}
  \includegraphics[width=0.9\textwidth]{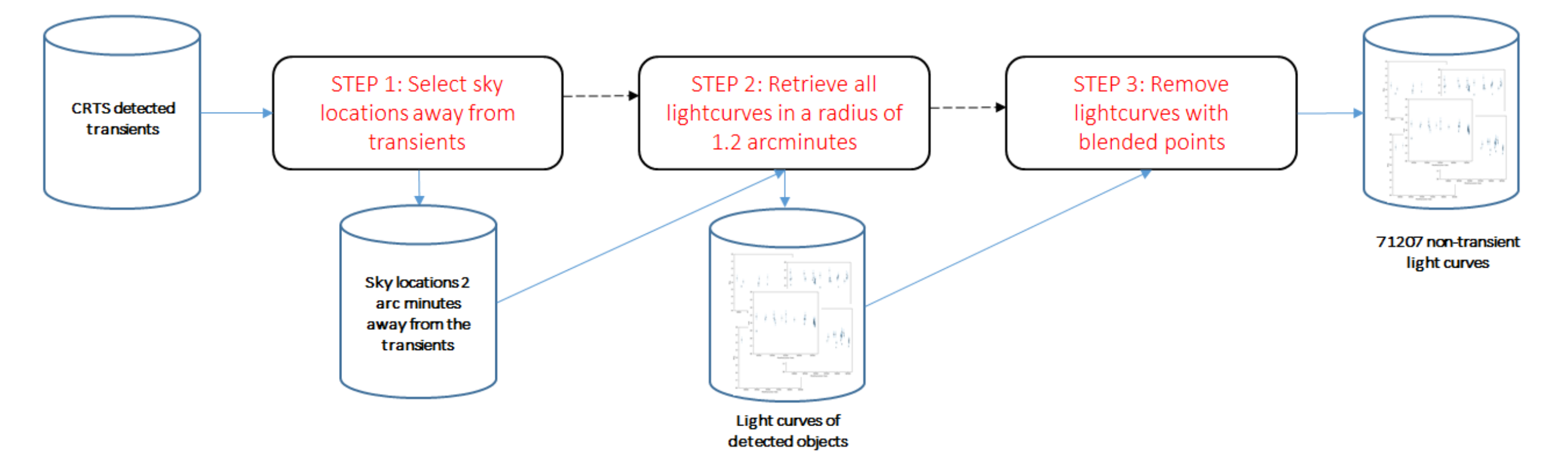}
\end{center}
  \caption{MANTRA Dataset Set Up: Lightcurve compilation for non-transients.}
  \label{fig:non-transients}
\end{figure*}

{
We also compile lightcurves for non-transients.
To do that we select sky locations
$2$ arc\-minutes away from the transients.
By construction the number of these locations is equal to the number of transients.
Then we query  \url{http://nesssi.cacr.caltech.edu/cgi-bin/getmulticonedb_release2.cgi}
to retrieve all lightcurves in a radius of $1.2$ arcminutes.
Each point in the retrieved lightcurves has a flag named {\texttt{blend}} indicating whether
the photometry for that source was blended with another source.
We only keep the lightcurves that do not have any point labeled as blended.
In this way we compile a total number of $71207$ non-transient lightcurves.
That is, we have approximately $15$ non-transient for every transient.
Figure \ref{fig:non-transients} illustrates this process.}

{For all lightcurves we compute a reduced $\chi^2$ statistic to quantify how different is 
the lightcurve from a constant lightcurve equal to the its average magnitude
\begin{equation}
    \chi^2_r = \frac{1}{N} \sum_{i=1}^{N}\left(\frac{m_i - \bar{m}}{\sigma_i}\right)^2,
    \label{eq:chi}
\end{equation}
where $N$ is the number of points in the lightcurve, $m_i$ is the magnitude of the 
$i$-th data point and $\sigma_i$ is its corresponding uncertainty and $\bar{m}$ is the average
magnitude over the lightcurve.
Not all non-transiet lightcurves have $\chi^2_r<1$ as one could expect from a statistically flat lightcurve.
Either instrumental, atmospheric or intrinsic variability can produce lightcurves with $\chi^2_r>1$. }

{In our case we find $25654$ light-curves with $\chi^2_r<1$ and $45553$ lightcurves with $\chi^2_r>1$. 
Lightcurves in the first case are called \emph{non-variable} and in the 
latter case \emph{variable}.
In the classification tests we present in this paper we only used non-variable non-transients.}

{We highlight that we are compiling the classification done by CRTS after they use additional photometric information, spectrometric follow-up, image processing and comparison with other catalogs \citep{2009ApJ...696..870D,2011BASI...39..387M}
to find a transient.
All the possible multiple instrumental and atmospheric effects that might produce variability and $\chi^{2}_r>1$ are naturally included in the database and cannot
be used a the sole evidence to define a transient.
On the same token, not all transients are trivial to detect, some have lightcurves with $\chi^{2}_r<1$, usually because they have a small number of measurements close to the faint magnitude limit of the observations.}

{This means that transient/non-transient classification from lightcurves is 
a complex task that surely requires more features than simply $\chi^2_r$.
Additional challenges in the classification of lightcurves 
include the inherent nature of transient events, which is reflected in different brightness behaviors, their evolution over time, and 
the nonuniform sampling of observations at sequential dates.}

\begin{figure*}
\begin{center}
  \includegraphics[width=0.45\textwidth]{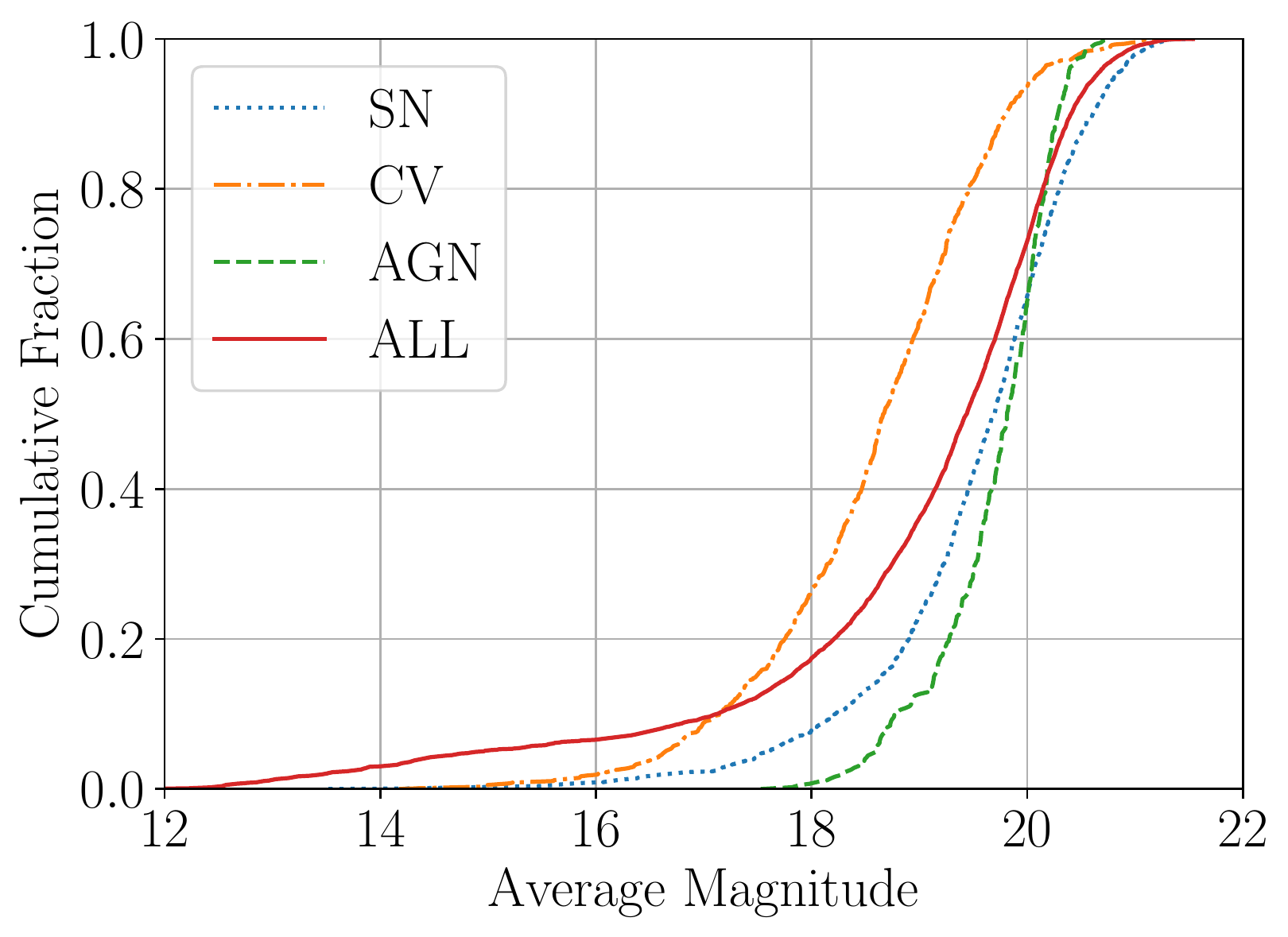}
  \includegraphics[width=0.45\textwidth]{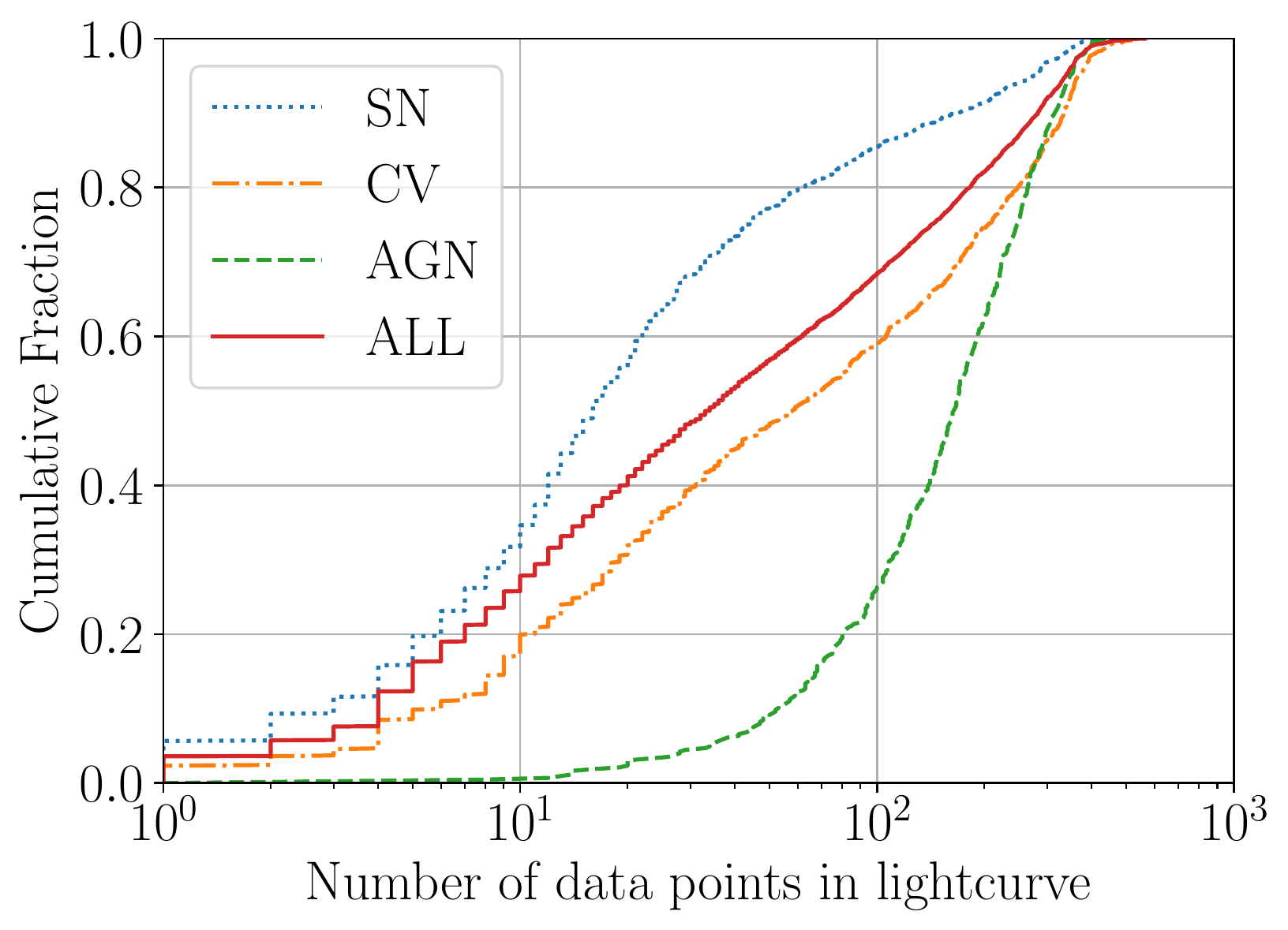}
\end{center}
  \caption{Cumulative number of lightcurves (expressed as a fraction)
    as a function of average magnitude (left) and number of data
    points in the lightcurve (right).
    This includes information for the three most representative
    classes (SN, CV, AGN) and the whole database (ALL).}
  \label{fig:cumulative}
\end{figure*}

\begin{figure*}
\begin{center}
  \includegraphics[width=0.6\textwidth]{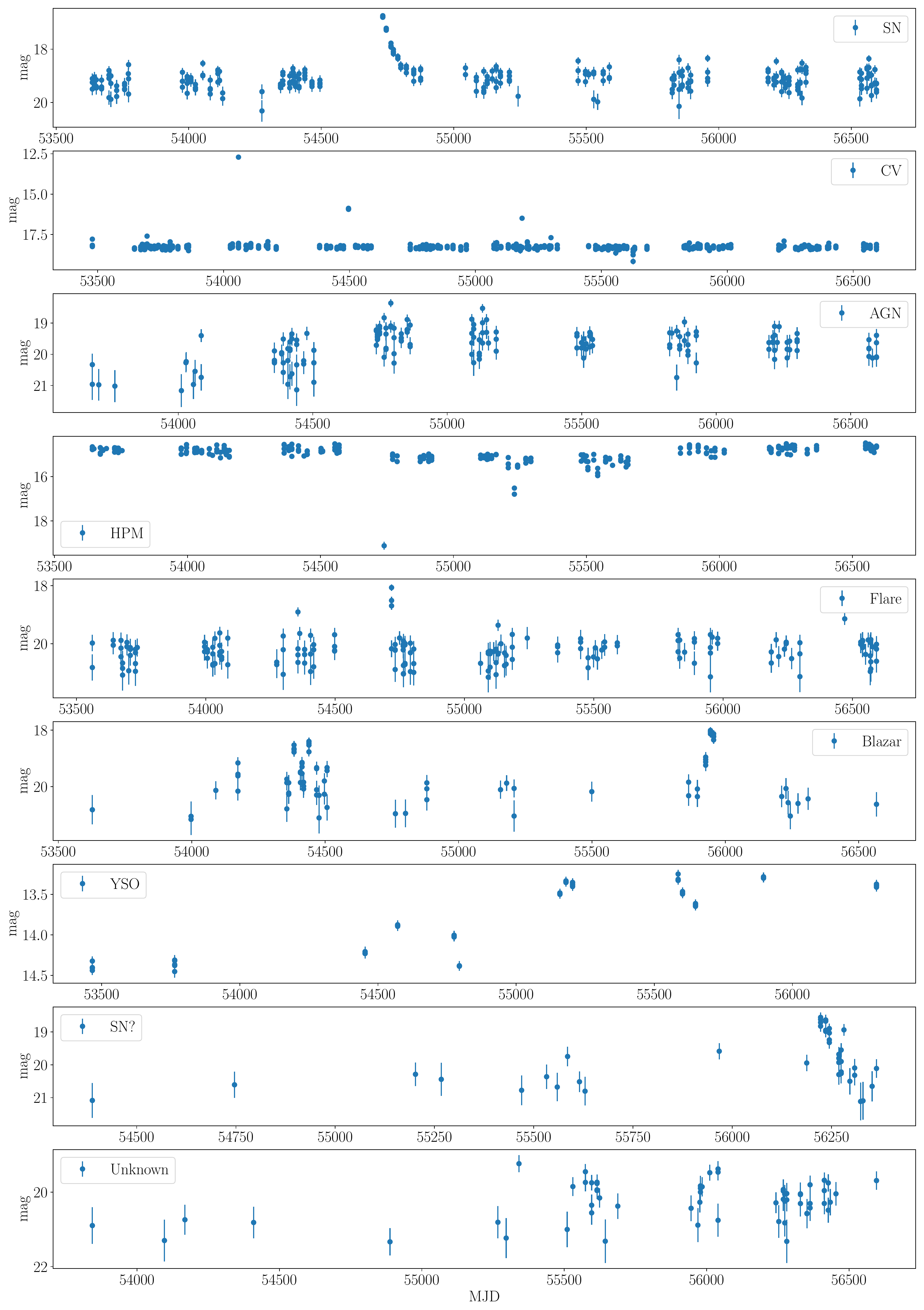}
\end{center}
  \caption{Randomly selected lightcurves for the most represented transient classes as compiled in MANTRA. The class of each sample is within the legend box.}
  \label{fig:examples_transient}
\end{figure*}

\begin{figure*}
\begin{center}
  \includegraphics[width=0.6\textwidth]{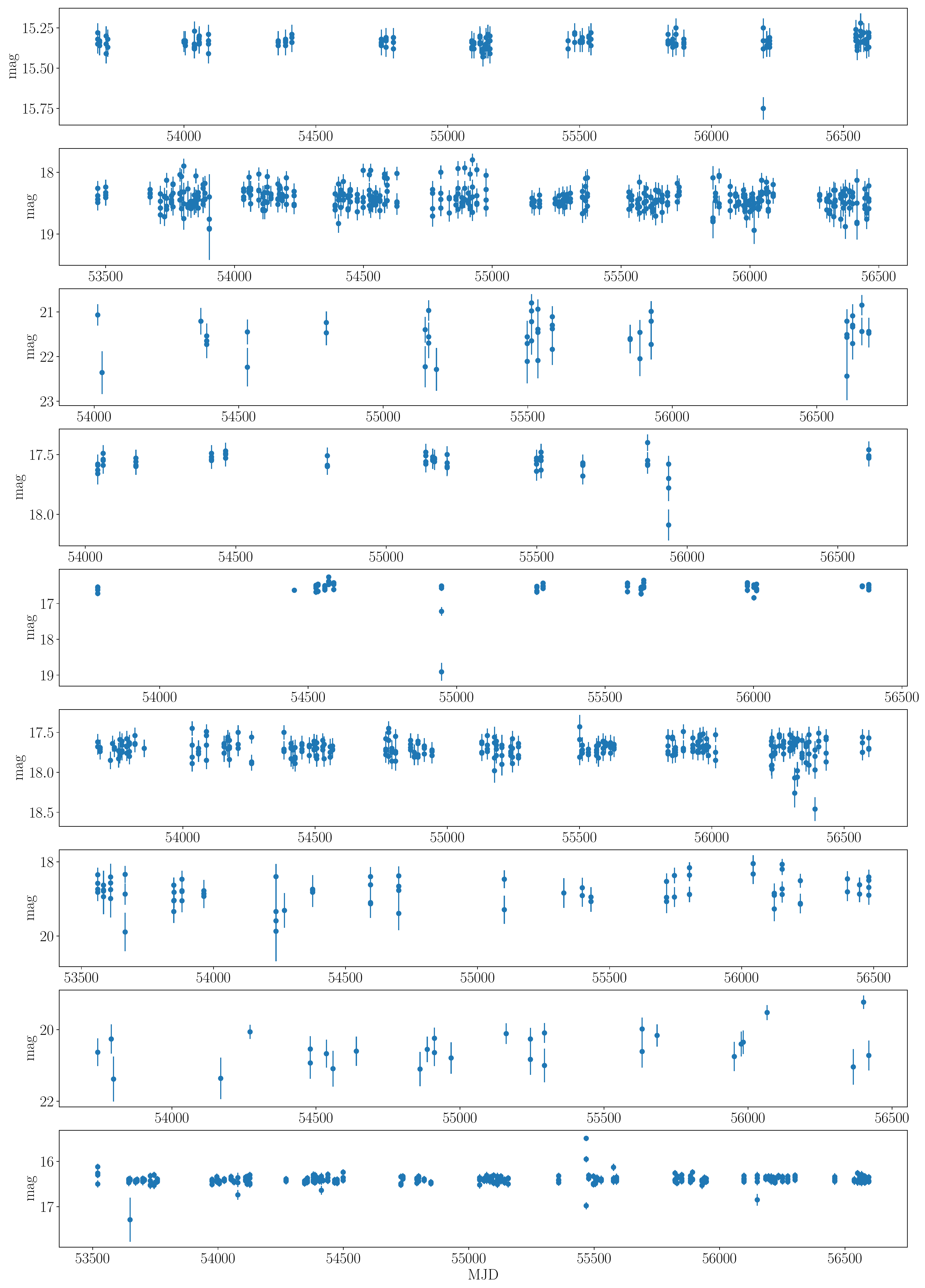}
\end{center}
  \caption{Randomly selected lightcurves for non-transient sources retrieved for MANTRA.}
  \label{fig:examples_non_transient}
\end{figure*}

Figure \ref{fig:cumulative} shows the number of lightcurves as a
function of average magnitude (left panel) and as a function of the
number of points in the lightcurve (right panel).
We show separately the whole data set and three most representative
classes: supernova, cataclysmic variables and active galactic nuclei. 
For these four sets, the median magnitude is in the range $18-20$. 
The number of points in the lightcurve has a larger variability.
The median for all the curves is close to 30, while for SN, CV and
AGN it is close to 15, 50 and 180, respectively. 
We provide sample lightcurves of the most represented transient classes 
and non-transient sources in Figure \ref{fig:examples_transient} and Figure \ref{fig:examples_non_transient}, respectively. 
The brightness evolution of non-transient sources is more stable over time, 
while transient objects present non-periodical changes at different time scales. 

\begin{table}
\centering
\begin{tabular}{c|c}
    \hline
    Class &  Object Count \\
    \hline
SN & 1723 \\
CV & 988 \\
HPM & 640 \\
AGN & 446 \\
SN? & 319 \\
Blazar & 243 \\
Unknown & 228 \\
Flare & 219 \\
AGN? & 138 \\
CV? & 77 \\
    \hline
\end{tabular}
\caption{Top 10 transient classes in the CRTS with their respective number of lightcurves.} 
\label{table:top_classes}
\end{table}

\subsection{Classification Tasks} \label{subsection_classification}
We study two classification tasks on the MANTRA dataset: 

\begin{itemize}
\item {Binary Classification}.
Using a balanced number of events from both classes in order 
to investigate the capability of distinguishing between transient
and non-transient sources.
\item{8-Class Classification}.
Using the unbalanced number of objects across classes to 
perform a classification into the following categories:
AGN, Blazar, CV, Flare, HPM, Other, SN and Non-Transient.
\end{itemize}

We evaluate both tasks using the metrics of a detection problem. 
For each class in the testing set, we report the maximum F1-Score 
that is defined as the harmonic mean of precision and recall. 
We construct Precision-Recall (PR) curves by setting different 
thresholds on the output probabilities of belonging to each class.

\section{Repository Description} 
\label{sec:repository}

The repository contains the lightcurves and Jupyter notebooks
to reproduce some of the Figures and Tables in this paper.
The repository can be found in \url{https://github.com/MachineLearningUniandes/MANTRA}. 
To date the repository has two main folders:
\begin{itemize}

\item \texttt{data/lightcurves}: 
contains the transient lightcurves (\path{transient_lightcurves.csv}),
the labels for the transients (\path{transient_labels.csv}), additional
information for each transient (\path{transient_info.csv}) and the lightcurves
for non-transient objects (eight different files
\path{nontransient_lightcurves_*.csv}). 
The first two files can be linked by unique transient IDs and
provided in the CRTS database.
\item \texttt{nb-explore}: includes a Jupyter notebook
  (\path{explore_light_curves.ipynb}) with examples on how to read
  and plot transient and non-transient lightcurves, extract the statistics in Table
  \ref{table:top_classes} and prepare the summary statistics in Figure
  \ref{fig:cumulative}. 
  Additional python files (\texttt{features.py},
  \texttt{helpers.py} and \texttt{inputs.py}) allow to read and perform simple operations on the CSV data files. 
\end{itemize}

\section{Machine Learning Methods} 
\label{sec:ml_tests}
\begin{figure*}
\begin{center}
  \includegraphics[width=0.9\textwidth]{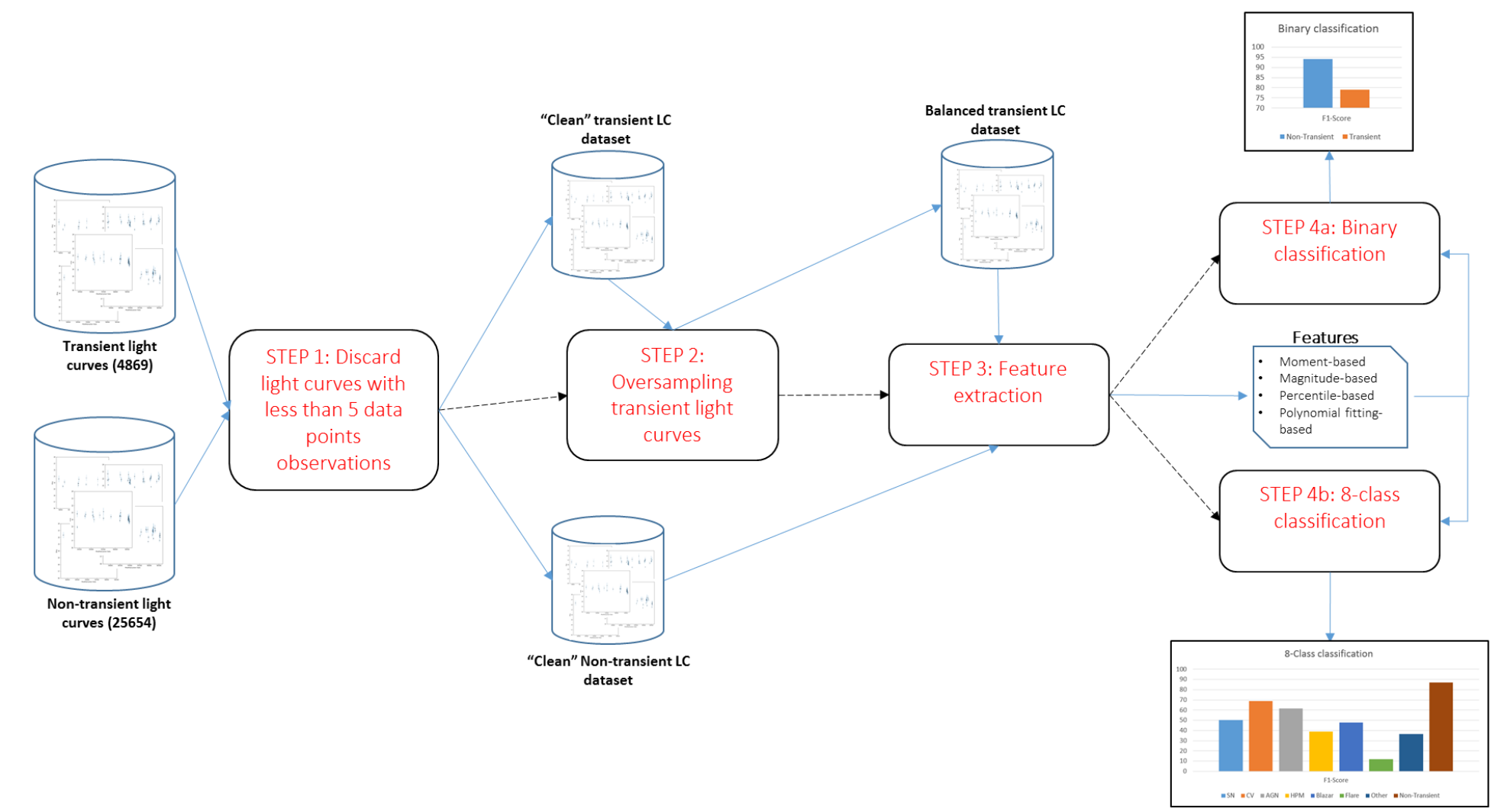}
\end{center}
  \caption{Overview of the Machine Learning process on the MANTRA
    dataset for the binary and 8-class classification tasks. We take
    the raw lightcurves as input, preprocess the data (step 1) and
    balance the classes for the training phase (step 2). 
    We extract the features (step 3) that are feed into the ML
    algorithms that perform the classification task (step 4).}
  \label{fig:ML}
\end{figure*}

Here we test baseline algorithms on the MANTRA dataset that can be
used as a reference for future work.
Figure \ref{fig:ML} shows an overview of our transient classification framework.
The main steps include and initial data split for non-transients,  feature extraction for all lightcurves and classification.

\subsection{Split on non-transient variability}

Our dataset includes non-transients with different degrees of variability as quantified by the $\chi^2_{r}$ statistic defined in Eq. (\ref{eq:chi}).
Here we report on the classification experiments using the non-transients with low variability ($\chi^2_r<1$), 
that is using 25654 lightcurves out of the total 71207 non-transients in the dataset we are presenting in this paper. These results are presented in great detail in Section \ref{sec:results}.
We perform separately the same classification experiments using only the high variability ($\chi^2_r\geq 1$) non-transients. Those results are summarized in the Appendix \ref{sec:results_variable}).

\subsection{Preprocessing and Feature Extraction}

We do not input directly the annotated lightcurves to the ML algorithms.
We perform a preprocessing stage as follows. 
First, we discard lightcurves with less than 5 data points observations
as they may not contain enough information to be classified correctly.
The reduction in the dataset is only performed for the purpose of the
classification tasks we present in this paper.
The public dataset includes all light curves.

Given that the number of lightcurves per class is imbalanced, 
in order to have the same number of instances for each class, we implement an
oversampling step by artificially generating multiple mock lightcurves 
from an observed one.
We generate a slightly different lightcurve from the observed lightcurve and 
then sample the observed magnitude from a Gaussian distribution
centered on the observational apparent magnitude with the magnitude's
error as the standard deviation. It is important to note that the oversampling was only done on the training set.  The test set was left unchanged.

Finally, we compute a standard set of features for each lightcurve. 
These features are scalars derived from statistical and model-specific
fitting techniques.
The first features (moment-based, magnitude-based and
percentile-based) were formally introduced in 
\cite{1101.1959}, and have been used in other studies \citep{1603.00882,disanto}.
We extend that list to include another set (polynomial fitting-based features) that explicitly take into account the time dependence of the lightcurves.

These groups of features are:

\begin{enumerate}
    
\item Moment-based features, which use the magnitude for each lightcurve.
  \begin{itemize}
  \item \texttt{beyond1std}: 
    Percentage of observations which are over or under one standard
    deviation from the weighted average. Each weight is calculated as
    the inverse of the corresponding observation's photometric error. 
  \item \texttt{kurtosis}: 
    The fourth moment of the data distribution. 
  \item \texttt{skew}: 
    Skewness. Third moment of the data distribution.
  \item \texttt{sk}:
    Small sample kurtosis.
  \item \texttt{std}:
    The standard deviation.
  \item \texttt{stetson\_j}:
    The Welch-Stetson J variability index
    \citep{1996PASP..108..851S}. A robust standard deviation. 
  \item \texttt{stetson\_k}:  The Welch-Stetson K variability index
    \citep{1996PASP..108..851S}. A robust kurtosis measure. 
  \end{itemize}
  
\item Features based on the magnitudes.
    \begin{itemize}
    \item \texttt{amp}: 
      The difference between the maximum and minimum magnitudes.
    \item \texttt{max\_slope}: 
      Maximum absolute slope between two consecutive observations.
    \item \texttt{mad}: 
      The median of the difference between magnitudes and the median
      magnitude. 
    \item \texttt{mbrp}: 
      The percentage of points within 10\% of the median magnitude.
    \item \texttt{pst}: 
      Percentage of all pairs of consecutive magnitude measurements that have positive slope.
    \item \texttt{pst\_last30}: 
      Percentage of the last 30 pairs of consecutive magnitudes that
      have a positive slope, minus percentage of the last 30 pairs of
      consecutive magnitudes with a negative slope. 
    
    \item \texttt{rcb}: Percentage of data points whose magnitude is below 1.5 mag of the median. 
    \item \texttt{ls}: Period of the peak frequency of the Lomb-Scargle periodogram (\cite{lombScargle})
    
    \end{itemize}

  \item Percentile-based features, which use the sorted flux distribution for
    each source. The flux is computed as $F = 10^{0.4 \mathrm{mag}}$. 
    We define $F_{n,m}$ as the difference between the $m$-th and $n$-the flux
    percentiles. 
    \begin{itemize}
    \item \texttt{p\_amp}: 
      Largest percentage difference between the absolute maximum
      magnitude and the median. 
    \item \texttt{pdfp}: 
      Ratio between $F_{5,95}$ and the median flux.
    \item \texttt{fpr20}: 
      Ratio $F_{40,60} / F_{5,95}$
    \item \texttt{fpr35}:
      Ratio $F_{32.5,67.5} / F_{5,95}$
    \item \texttt{fpr50}: 
      Ratio $F_{25,75} / F_{5,95}$
    \item \texttt{fpr65}: 
      Ratio $F_{17.5,82.5} / F_{5,95}$
    \item \texttt{fpr80}: 
      Ratio $F_{10,90} / F_{5,95}$
    \end{itemize}
    
  \item Polynomial Fitting-based features, which are the coefficients of
    multi-level terms in a polynomial curve fitting. This is a new set
    of features proposed in this paper. 
    \texttt{Polyn\_Tm} indicates the coefficient of the term of order
    \texttt{m} in a fit to a polynomial of order \texttt{n}.
    \begin{itemize}
        \item \texttt{Poly1\_T1}.
        \item \texttt{Poly2\_T1}.
        \item \texttt{Poly2\_T2}.
        \item \texttt{Poly3\_T1}.
        \item \texttt{Poly3\_T2}.
        \item \texttt{Poly3\_T3}.
        \item \texttt{Poly4\_T1}.
        \item \texttt{Poly4\_T2}.
        \item \texttt{Poly4\_T3}.
        \item \texttt{Poly4\_T4}.
    \end{itemize}    
\end{enumerate}

\subsection{ML algorithms}

We conduct experiments with three widely used families of supervised classification 
algorithms \citep{skysurvey, disanto}: Neural Networks (NNs),
Random Forests (RFs) and Support Vector Machines (SVMs).  

These algorithms are popular in published studies and are efficient for low dimensional feature datasets as is our case. 
We use SciKit-Learn \citep{1201.0490} Python's implementation of random forests and support vector machines.
Details on the inner workings of these machine learning models can be found in \cite{9780387848570}. 

We use the pytorch library for python for the development of the
linear Neural Networks.
It consists of a series of fully connected layers that map the
features to the corresponding number of classes. At each layer, a 1d batch normalization is implemented followed by a rectified linear unit (ReLU) activation function.
The final layer calls a softmax activation function to
transform the numerical values to class probabilities. 

Note that for SVMs, the features were normalized to have zero mean and unit variance. The test set was normalized with respect to the training set. The hyperparameters explored for each algorithm are the
following. 

\begin{itemize}
\item Neural Networks:
\begin{itemize}
\item Learning Rate: $\{0.1,0.01,0.001,0.0001\}$
\item Hidden Layer Sizes: Single , double or triple layers with $500$ nodes each.
\end{itemize}

\item Random Forest:
\begin{itemize}
    \item Number of Estimators: $200$ or $700$.
    \item Number of features considered: Square root or $log_2$ of the total number of features. 
\end{itemize}

\item Support Vector Machines:
\begin{itemize}
    \item Kernels: Radial Basis Function (RBF), linear or sigmoid.
    \item Kernel Coefficient ($\gamma$):  
      $\{0.125, 2, 32\}$
    \item Error Penalty (\textit{C}): $\{0.125, 2, 32\}$
\end{itemize}
\end{itemize}

\subsection{Validation} \label{subsection_importances}

We split the input lightcurves into training and testing in a $75:25$
ratio respectively, class by class. 
For the random forests and the SVM, we use a grid search over the
hyperparameter combinations with a 2-fold cross-validation over the
training set to determine the best hyperparameters.  
For the neural networks, at each epoch, the network is evaluated on
the test data.

\subsection{Results}
\label{sec:results}

Table \ref{table:all-avg-results} shows the average class precision,
recall and F1-measure for each of the classification tasks and
algorithms listed above.  

    We also compare our classification with that of \cite{disanto}. Given that they do not report their F1 scores nor the confusion matrices, we perform the training on our own using their features. 
    The results can be found in Table \ref{disantoComparison}. We find that our results outperform their methodology by $0.87\%$ and $2.87\%$ on the F1 score for the binary and 8-class classification tasks respectively. We consider that modeling the temporal dimension by including the polynomial features is responsible for this improvement. 
    Their relative importance in Figures \ref{binaryImportance_nonvar} and \ref{8classImportance_nonvar} supports this hypothesis.

\begin{table}
\centering
\begin{tabular}{ccccc}
\hline
\multicolumn{1}{l}{\textbf{Case}} & \textbf{Classifier} & \textbf{Precision} & \textbf{Recall} & \textbf{F1-score} \\ \hline \hline
\multirow{3}{*}{Binary}                 & RF                  & \textbf{96.35}    & \textbf{96.15} & \textbf{96.25}   \\
                                        & SVM                 & 95.33             & 93.94          & 94.61             \\
                                        & NN                  & 84.61              & 84.81           & 84.71             \\ \hline
\multirow{3}{*}{8 Class}                & RF                  & \textbf{49.12}     & \textbf{69.60}  & \textbf{52.79}             \\
                                        & SVM                 & 33.62              & 60.34           & 37.59             \\
                                        & NN                  & 24.14     & 60.21           & 29.22   \\ \hline
\end{tabular}%
\caption{Average precision, recall and F1-score accross all classes
  for each algorithm and classification task. Best results per metric
  per classification task are in bold.} 
\label{table:all-avg-results}
\end{table}

\begin{figure}
\begin{center}
  \includegraphics[width=0.55\textwidth]{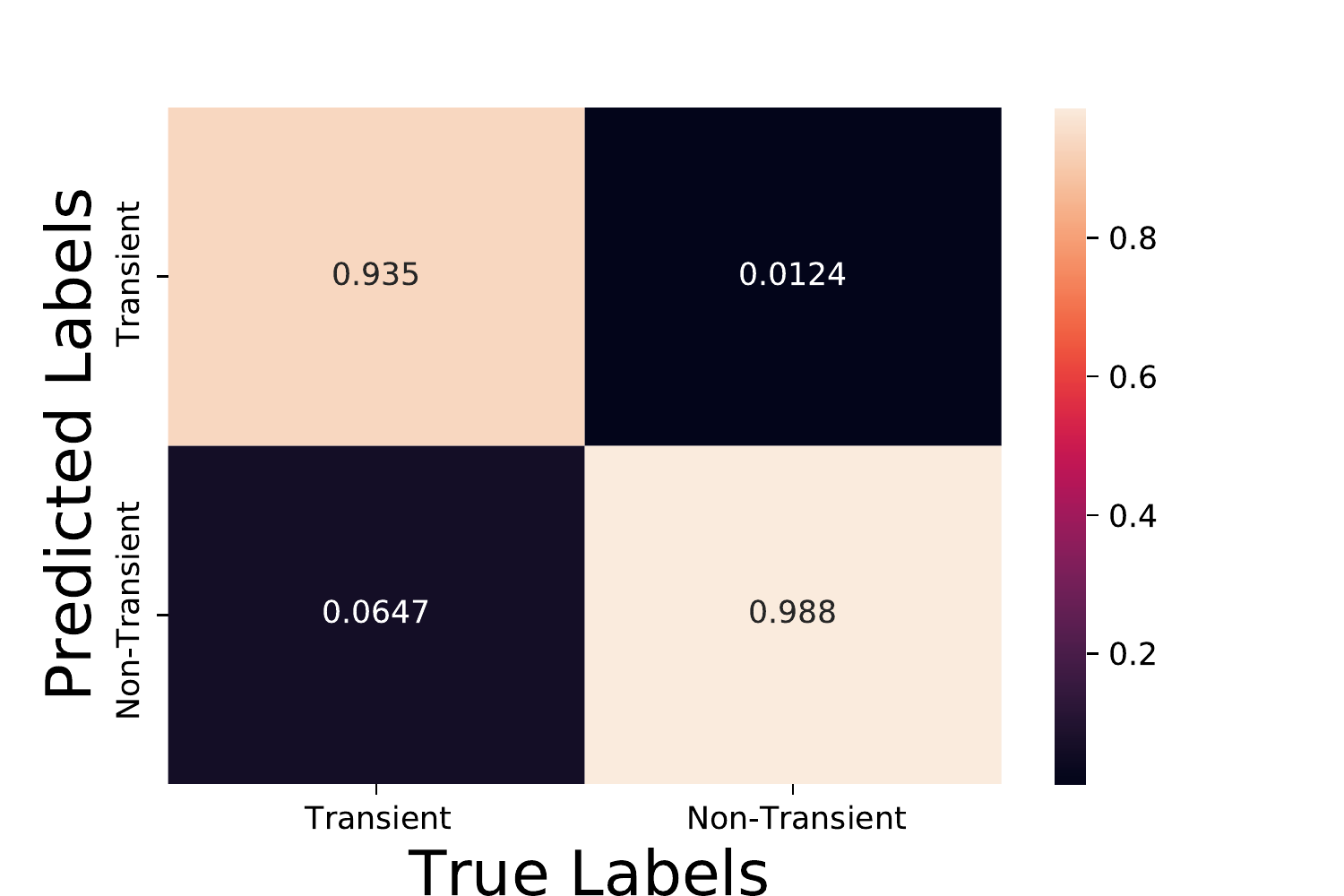}
\end{center}
  \caption{Confusion Matrix for the best performing model in the
    Binary task. Rows represent prediction and columns the ground
    truth.} 
  \label{fig:normalizedBinaryCM_nonvar}
\end{figure}

\begin{figure*}
\begin{center}
\includegraphics[width=1.1\textwidth]{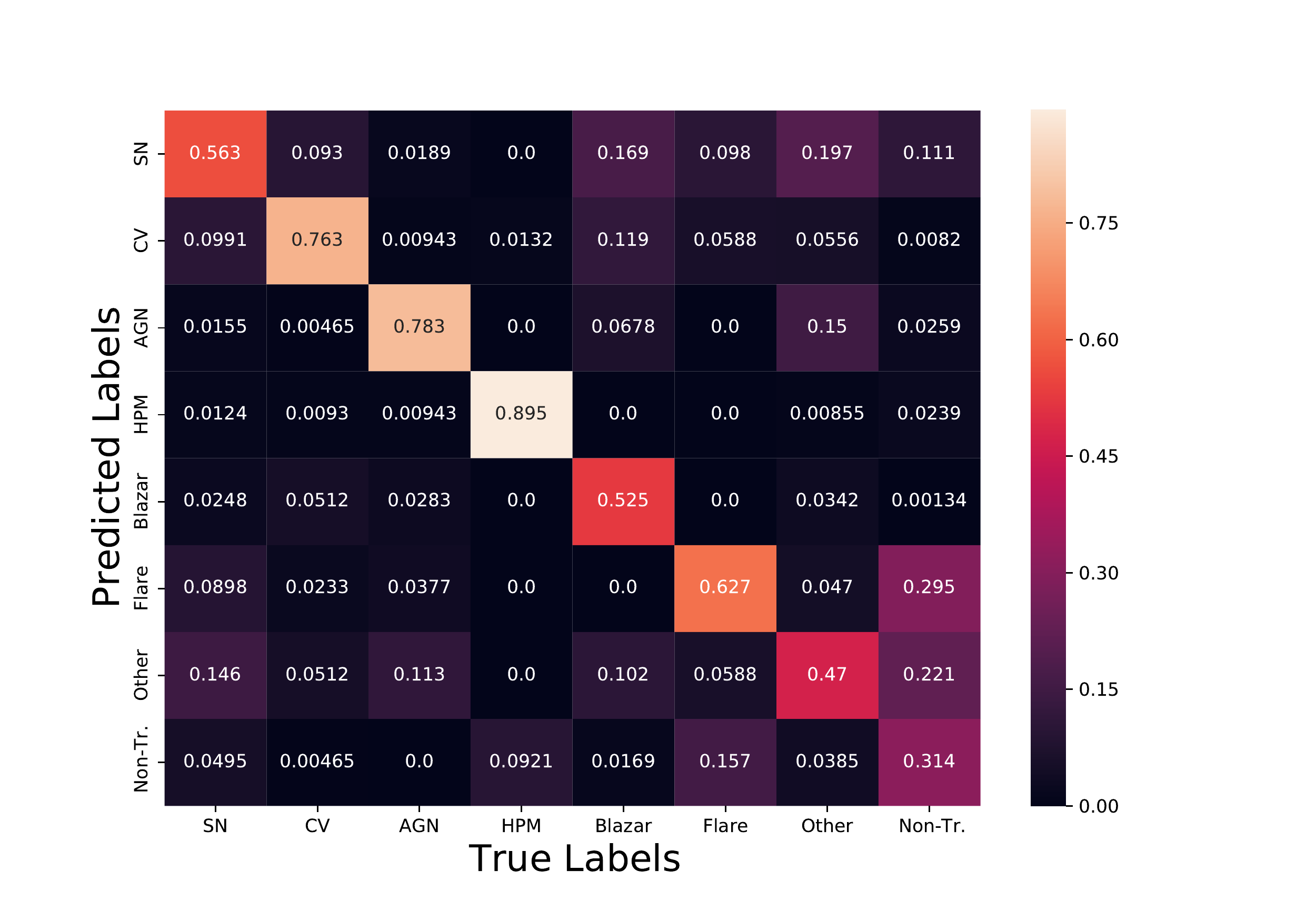}
\end{center}
  \caption{Confusion Matrix for the best performing model in the 8-class
  task. The classes follow the abbreviations in Table
  \ref{Overall-Scores-8-Class-Regular}. Rows represent the
  predictions, and columns the ground truth.} 
  \label{fig:normalized8ClassCM_nonvar}
\end{figure*} 

\subsubsection{Binary Classification} 
\label{Results-Binary}

The best algorithm in this task is RFs with an average F1-Score of
$96.25\%$.   
SVMs are the second best-performing model with a F1-Score of $94.61\%$. 
NNs are ranked third with an F1-Score of $84.71\%$.

Figure \ref{fig:normalizedBinaryCM_nonvar} shows the confusion matrix of the best
performing algorithm. These results suggest that in an imbalanced set
up, non-transient sources are better classified while transients are
more difficult, showing a difference of about  5 points in the
percentage of correct classifications.
This difference in performance could be attributed to the intra-class
variation within the transients.

\begin{figure*}
\begin{center}
\makebox[\textwidth][c]{\includegraphics[width=1.1\textwidth]{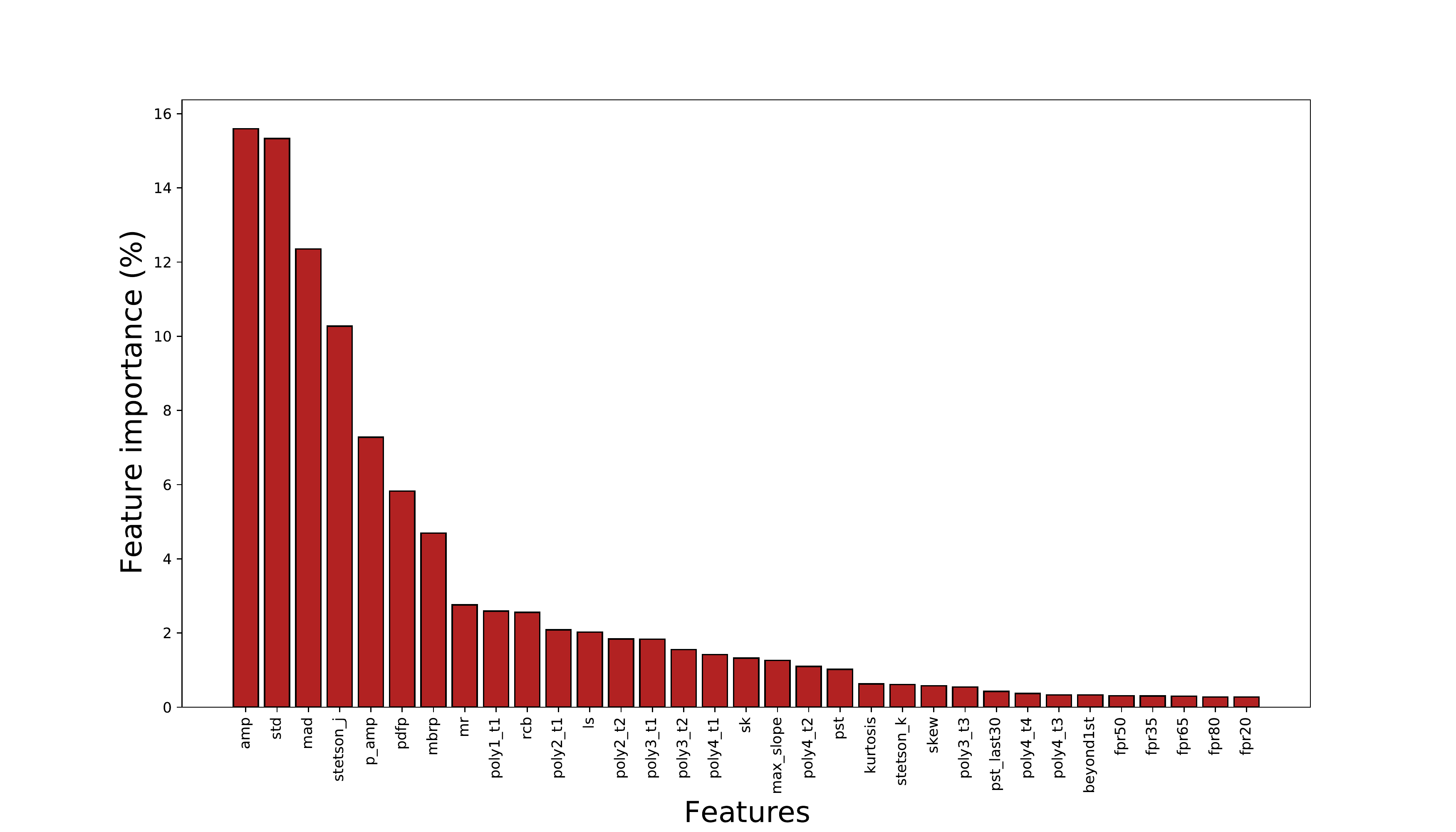}}%
\end{center}
  \caption{Feature importance for the binary classification task.} 
  \label{binaryImportance_nonvar}
\end{figure*}

The feature importance list for this problem can be seen in Figure \ref{binaryImportance_nonvar}. We find that the order of the features is akin to that found by \cite{disanto}. Their top 5 features ($poly1\_t1$, $std$, $ampl$, $l$ and  $skew$) are in our top 10 most important features for classification excluding $ls$ (the lomb-scargle periodogram) which was not included. In general, the first term of the polynomials is significantly more important than the following terms.

\subsubsection{Eight-Class Classification}

For this task, RF is again the best classifier.
The best F1-Score is $52.79\%$. 
SVMs are the second best model with an F1-Score of $37.59\%$, while
NNs are the worst-performing model only achieving an average F1-Score
of $29.22\%$.
Table \ref{Overall-Scores-8-Class-Regular} summarizes the results for 
individual classes and Figure  \ref{fig:normalized8ClassCM_nonvar} presents 
the confusion matrix for the RF.

The two classes with highest F1-Score are non-transient ($96.83\%$) and
CV ($75.23\%$).  
The recall decreases for the non-transient class in comparison to the binary experiment, 
meaning that the algorithm misclassified some instances that belong to non-transient class among transient classes. 
However, transient sources are not commonly confused with non-transient ones. 
The worst performing classes are Flare, Other and HPM, with F1-Scores in the 
range 16\% - 37\%. 
A possible reason for that is that flaring events are rare and
are short lived (lasting tens of minutes) and would then typically span few 
datapoints in the lightcurve.
It is worth noting that the less frequent classes present a lower performance, 
such as Flare and HPM. 
Even though the most frequent classes are more easily identified, 
the "other" type class has a low F1-score due to the diverse nature of sources assigned to this category. 

\begin{figure*}
\begin{center}
\makebox[\textwidth][c]{\includegraphics[width=1.1\textwidth]{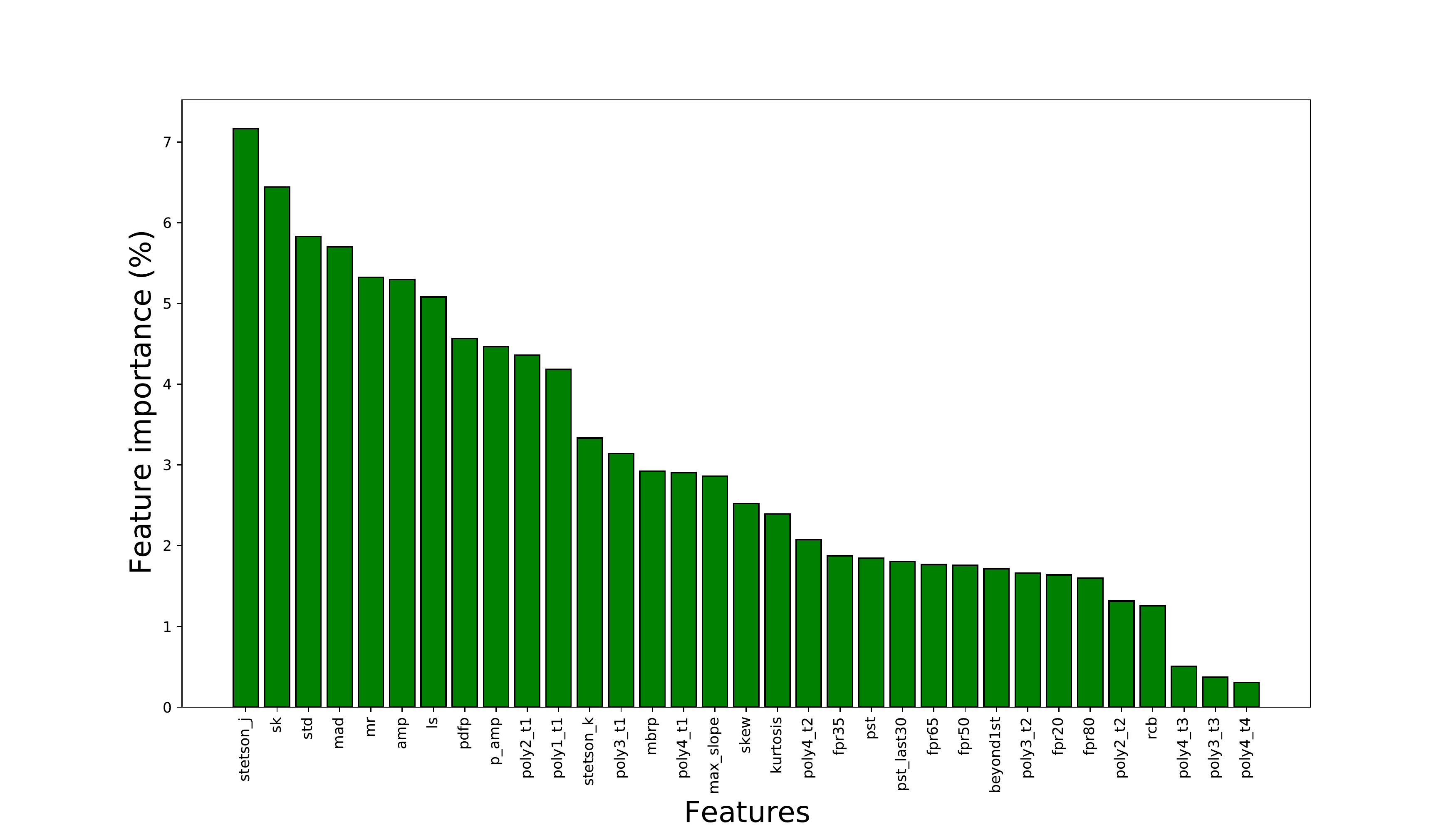}}%
\end{center}
  \caption{Feature importance for the 8 class classification task.} 
  \label{8classImportance_nonvar}
\end{figure*}

The feature importance list can be found in figure \ref{8classImportance_nonvar}. 
Even though some features have been displaced, the general order with respect to Figure \ref{binaryImportance_nonvar} has been preserved.

\begin{table}
\centering
\begin{tabular}{ccccc}
\hline
\textbf{Class} & \textbf{Precision} & \textbf{Recall} & \textbf{F1-score}& \textbf{Cover} \\\hline \hline
SN        & 52.91 & 56.35 & 54.57 & 323   \\ \hline
CV        & 74.21 & 76.28 & 75.23 & 215   \\ \hline
AGN       & 63.85 & 78.30 & 70.34 & 106   \\ \hline
HPM       & 9.26  & 89.47 & 16.79 & 76    \\ \hline
Blazar    & 50.82 & 52.54 & 51.67 & 59    \\ \hline
Flare     & 11.99 & 62.75 & 20.13 & 51    \\ \hline
Other     & 30.14 & 47.01 & 36.73 & 234   \\ \hline
Non-Tr.   & 99.76 & 94.07 & 96.83 & 18556 \\ \hline
avg/total & 49.12 & 69.60 & 52.79 & 19620 \\ \hline

\end{tabular}%
\caption{Precision, Recall and F1-Score for the 8-Class Classification Task.}
\label{Overall-Scores-8-Class-Regular}
\end{table}

\begin{table}[]
\centering
\begin{tabular}{ccccc}
\hline
\textbf{Task}           & \textbf{Method} & \textbf{Precision} & \textbf{Recall} & \textbf{F1} \\ \hline \hline
\multirow{2}{*}{Binary} & D'Isanto & 95.92          & 94.86          & 95.38           \\
                        & Ours     & \textbf{96.35} & \textbf{96.15} & \textbf{96.25}    \\ \hline
\multirow{2}{*}{8Class} & D'Isanto & 46.55         & 66.76          & 49.92            \\
                        & Ours     & \textbf{49.12} & \textbf{69.60} & \textbf{52.79}   \\ \hline
\end{tabular}
\caption{Average precision, recall and F1 score for each of the classification tasks. In bold are the best results for each task.}
\label{disantoComparison}
\end{table}

\section{Conclusions}
\label{sec:conclusions}

The scope of forthcoming large astronomical synoptic surveys 
motivates the development and exploration of automatized ways to
detect transient sources. 
Such developments require observational datasets to train and test new
algorithms.  
Making these datasets public and easy to access has the potential
to open the field to a larger number of contributors.
With these goals in mind, in this paper we presented a compilation
based on data from the Catalina Real-Time Transient Survey.
The dataset has  $4869$ transient and $71207$ non-transient
lightcurves.
The dataset is publicly available at
\url{https://github.com/MachineLearningUniandes/MANTRA}.   

We illustrated how to use this database by extracting 
characteristic features to use them as input to train three different
machine learning algorithms (Random Forests, Neural Networks and
Support Vector Machines) for classification tasks.
The features extracted from lightcurves were either statistical
descriptors of the observations, or polynomial curve fitting
coefficients applied to the lightcurves.   
Overall, the best classifier for all tasks was the Random Forest.
Neural Networks showed the worst performance in the binary
classification task and the second best in the 8-class classification task. 

Certainly other classification algorithms could be used on the MANTRA
lightcurves. Our purpose here was not a thorough analysis of machine
learning algorithms. Our focus was the compilation of the data into
a format that could easily be used for different projects that could
be addressed with MANTRA. 
For instance the supernovae detection problem could be studied as a
classification problem between SN and the other classes by using incomplete
light curves, mimicking the process of observations that extend the light curve as a survey progresses.
Considering extremely unbalanced classes could also be addressed
with the dataset we are presenting by simply using less samples for
the SN class and keeping a large numbers for the non-transient class.  

In a second paper we will present the second part of MANTRA.
It corresponds to almost one million images from the CRTS.
These will be tested using state-of-the art deep learning techniques
for transient classification.  

\section*{Acknowledgements}

We thank Andrew Drake for sharing with us the CRTS Transient dataset
used in this project.  
We acknowledge funding from Universidad de los Andes.
We also thank contributors and collaborators of the SciKit-Learn,
Jupyter Notebooks and Pandas Python libraries.  

CRTS and CSDR2 are supported by the U.S.~National Science 
Foundation under grant NSF grants AST-1313422, AST-1413600, and 
AST-1518308.  The CSS survey is funded by the National Aeronautics
and Space Administration under Grant No. NNG05GF22G issued through
the Science Mission Directorate Near-Earth Objects Observations Program.

\appendix

\section{Results on the non-transients with high variability}\label{sec:results_variable}
For completeness, we applied the same tasks on the non-transients with high variability ($\chi^2_r\geq 1$). The main results are in Table \ref{table:all-avg-results_appendix}.

\begin{table}[h]
\centering
\begin{tabular}{ccccc}
\hline
\multicolumn{1}{l}{\textbf{Case}} & \textbf{Classifier} & \textbf{Precision} & \textbf{Recall} & \textbf{F1-score} \\ \hline \hline
\multirow{3}{*}{Binary}                 & RF                  & \textbf{85.26}    & \textbf{88.84} & \textbf{86.93}   \\
                                        & NN                  & 84.50              & 84.44           & 84.47          \\ 
                                        & SVM                 & 80.34             & 82.96          & 82.79             \\ \hline
\multirow{3}{*}{8 Class}                & RF                  & \textbf{27.50}     & \textbf{65.57}  & \textbf{33.48}             \\
                                        & NN                  & 24.62     & 59.75           & 29.49   \\
                                        & SVM                 & 18.86              & 56.67           & 21.31             \\ \hline                                        
\end{tabular}%
\caption{Average precision, recall and F1-score across all classes
  for each algorithm and classification task. Best results per metric
  per classification task are in bold.} 
\label{table:all-avg-results_appendix}
\end{table}

\subsection{Binary classification}

\begin{figure}[H]
\begin{center}
  \includegraphics[width=0.55\textwidth]{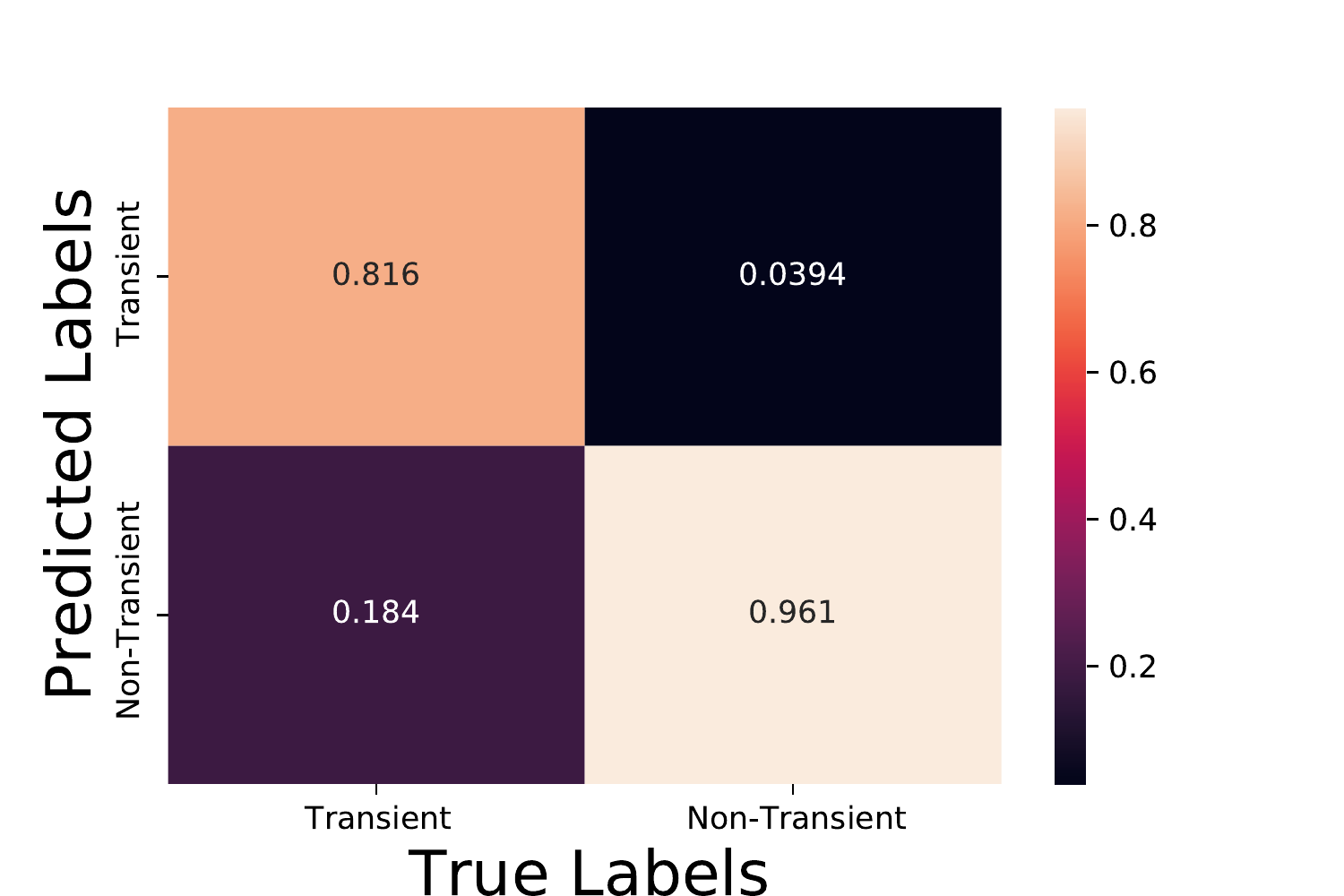}
\end{center}
  \caption{Confusion Matrix for the best performing model in the Binary task. Rows represent prediction and columns the ground
    truth.} 
  \label{fig:normalizedBinaryCM_appendix}
\end{figure}

\begin{figure}[H]
\begin{center}
 \includegraphics[width=1\textwidth]{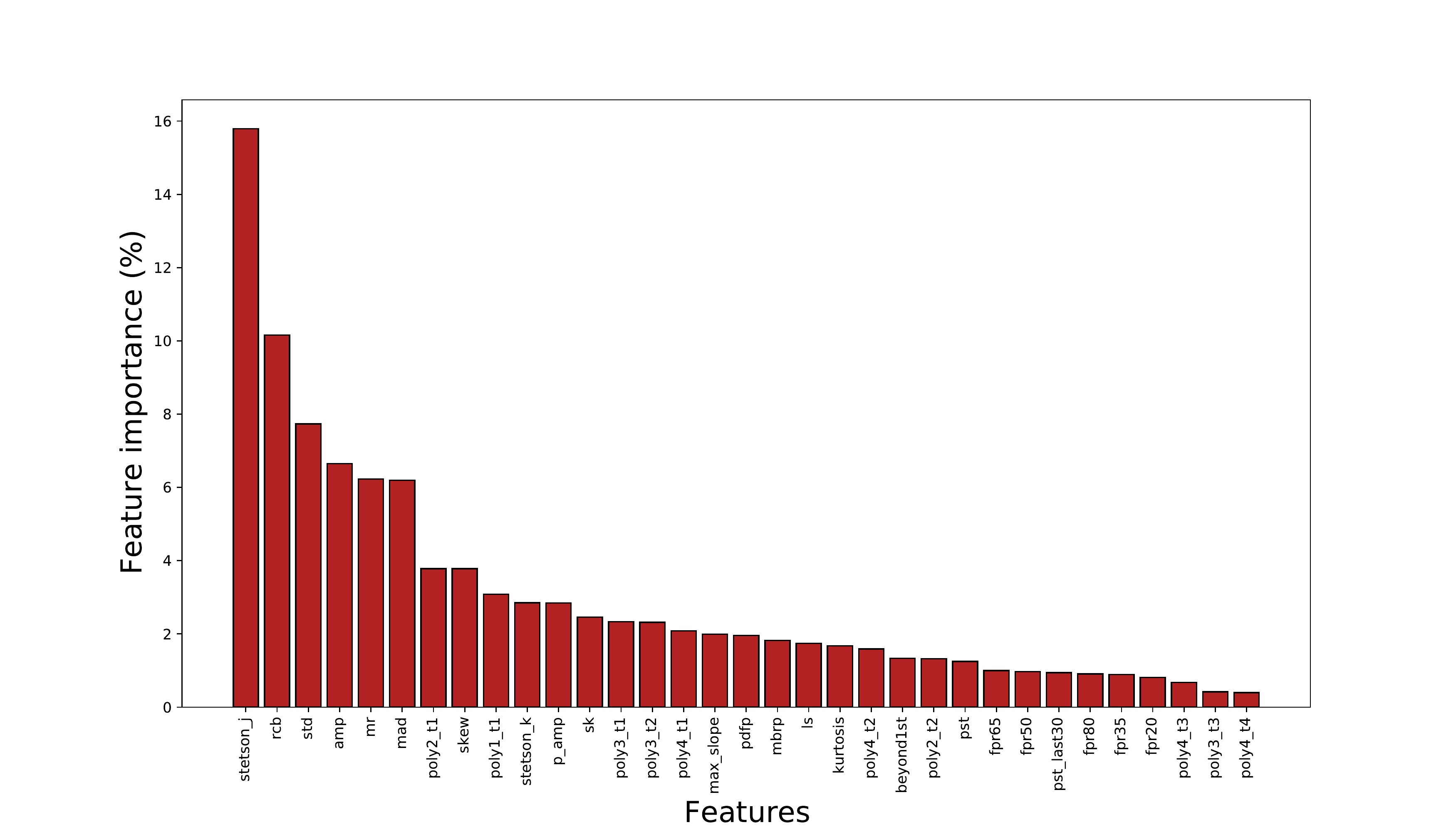}
\end{center}
  \caption{Feature importance for the binary classification task.} 
  \label{binaryImportance_appendix}
\end{figure}

 The best algorithm in the binary task is RFs with an average F1-Score of 86.93\%.  NN are the second best-performing model with a F1-Score of 84.47\%. SVM are ranked third with an F1-Score of 82.79\%. The Figure \ref{fig:normalizedBinaryCM_appendix} shows the confusion matrix of the best performing algorithm. The Figure \ref{binaryImportance_appendix} shows the feature importance for this problem.

\subsection{Eight-class classification}

\begin{figure}[H]
\begin{center}
\includegraphics[width=1\textwidth]{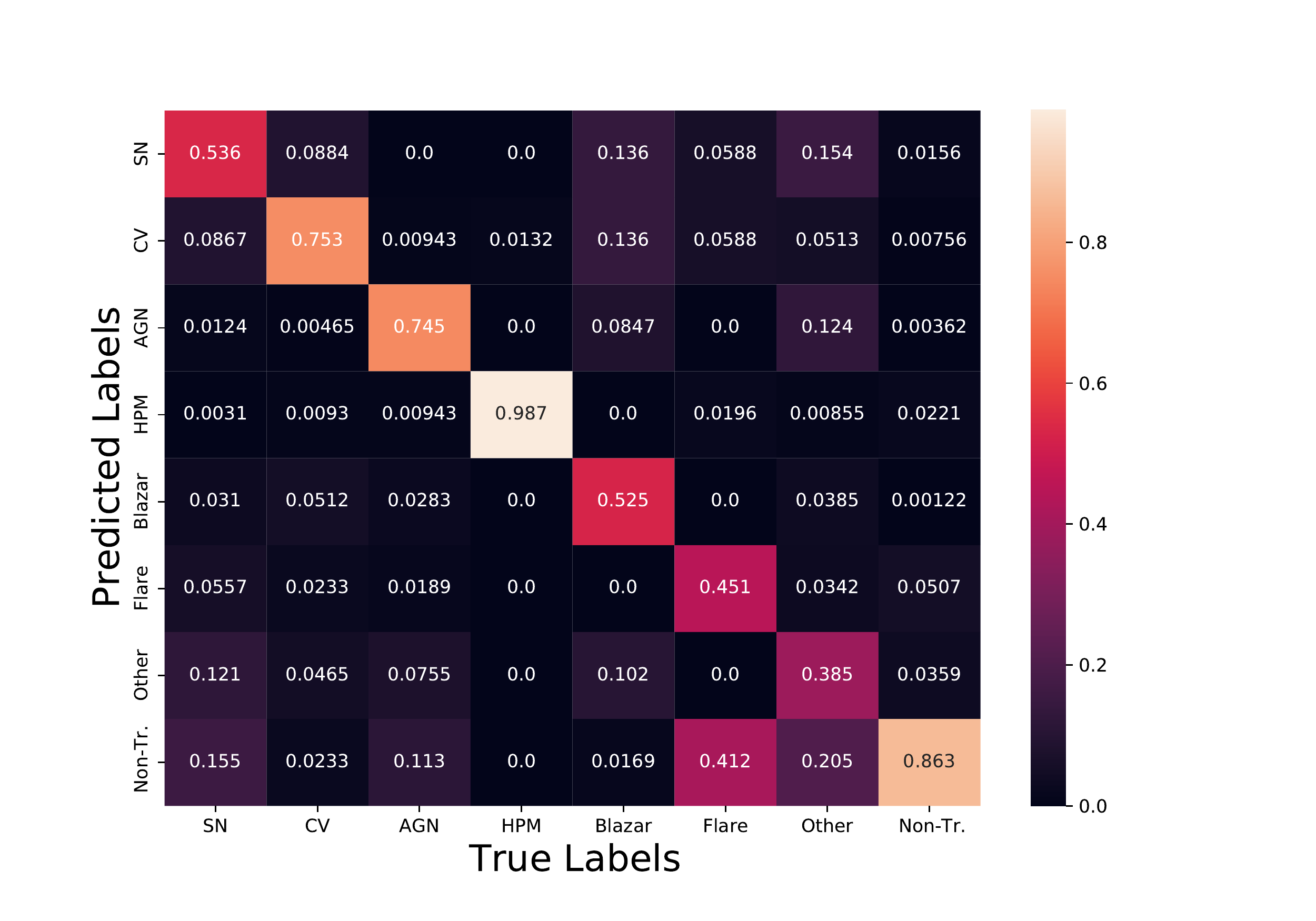}
\end{center}
  \caption{Confusion Matrix for the best performing model in the 8-class
  task. The classes follow the abbreviations in Table
  \ref{Overall-Scores-8-Class-Regular}. Rows represent the predictions, and columns the ground truth.} 
  \label{fig:normalized8ClassCM_appendix}
\end{figure}

\begin{figure}[H]
\begin{center}
 \includegraphics[width=1\textwidth]{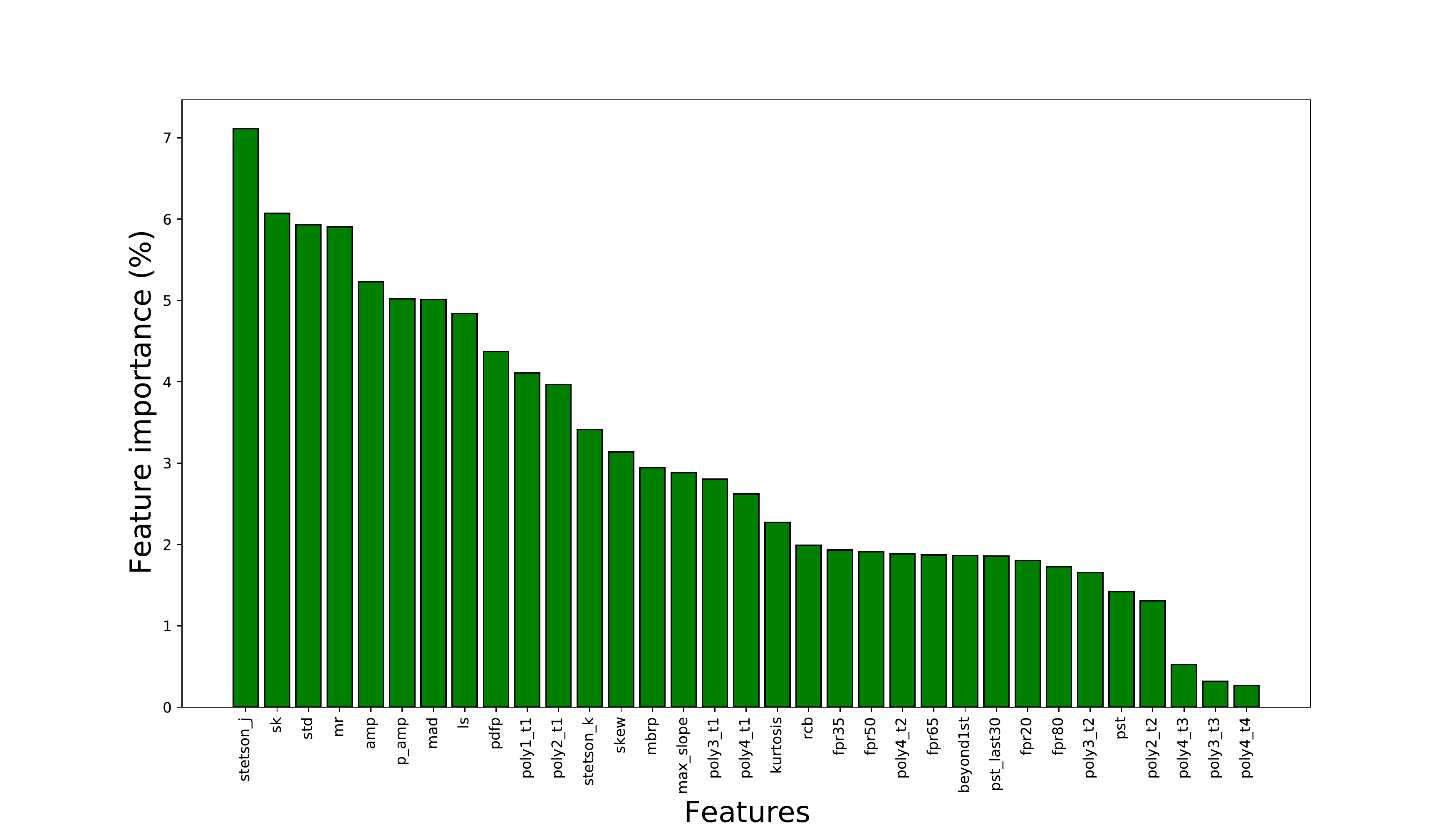}
\end{center}
  \caption{Feature importance for the 8 class classification task.} 
  \label{8classImportance_appendix}
\end{figure}

The best algorithm in the eight-class task also is RFs with an average F1-Score of 33.48\%.  SVMs are the second best-performing model with a F1-Score of 21.31\%. NNs are ranked third with an F1-Score of 29.49\%. The Table \ref{Overall-Scores-8-Class-Regular_appendix} summarizes the results for individual classes and the Figure \ref{fig:normalized8ClassCM_appendix} shows the confusion matrix for the best performing algorithm. The Figure \ref{8classImportance_appendix} shows the feature importance for this problem.

\begin{table}[H]
\centering
\begin{tabular}{ccccc}
\hline
\textbf{Class} & \textbf{Precision} & \textbf{Recall} & \textbf{F1-score}& \textbf{Cover} \\\hline \hline
SN        & 19.48 & 53.56 & 28.57 & 323   \\\hline
CV        & 30.57 & 75.35 & 43.49 & 215   \\\hline
AGN       & 29.37 & 74.53 & 42.13 & 106   \\\hline
HPM       & 7.46  & 98.68 & 13.88 & 76    \\\hline
Blazar    & 26.96 & 52.54 & 35.63 & 59    \\\hline
Flare     & 1.06  & 45.10 & 2.07  & 51    \\\hline
Other     & 5.45  & 38.46 & 9.55  & 234   \\\hline
Non-Tr.   & 99.62 & 86.33 & 92.50 & 41694 \\\hline
avg/total & 27.50 & 65.57 & 33.48 & 42758 \\ \hline
\end{tabular}
\caption{Precision, Recall and F1-Score for the 8-Class Classification Task.}
\label{Overall-Scores-8-Class-Regular_appendix}
\end{table}

\bibliographystyle{aasjournal}

\end{document}